# Scheduling Garbage Collection for Energy Efficiency on Asymmetric Multicore Processors

Marina Shimchenko[a] , Erik Österlund[b] , and Tobias Wrigstad[a]

a   Uppsala University, Sweden
b   Oracle, Sweden

**Abstract**   The growing concern for energy efficiency in the Information and Communication Technology (ICT) sector has prompted the exploration of resource management techniques. While hardware architectures, such as single-ISA asymmetric multicore processors (AMP), offer potential energy savings, there is still untapped potential for software optimizations. This paper aims to bridge this gap by investigating the scheduling of garbage collection (GC) activities on a heterogeneous architecture with both performance cores ("p-cores") and energy cores ("e-cores") to achieve energy savings.

Our study focuses on the concurrent ZGC collector in the context of Java Virtual Machines (JVM), as the energy aspect is not well studied in the context of latency-sensitive Java workloads. By comparing the energy efficiency, performance, latency, and memory utilization of executing GC on p-cores versus e-cores, we present compelling findings.

We demonstrate that scheduling GC work on e-cores overall leads to $\approx 3\%$ energy savings without performance and mean latency degradation while requiring no additional effort from developers. Overall energy reduction can increase to $5.3\% \pm .0225$ by tuning the number of e-cores (still not changing the program!).

Our findings highlight the practicality and benefits of scheduling GC on e-cores, showcasing the potential for energy savings in heterogeneous architectures running Java workloads while meeting critical latency requirements. Our research contributes to the ongoing efforts toward achieving a more sustainable and efficient ICT sector.

## The Art, Science, and Engineering of Programming

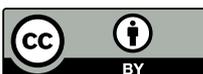

## 1 Introduction

The Information and Communication Technology (ICT) sector uses ≈3% (805TWh) of the global electricity [16] and is predicted to continue to grow [3, 7]. It is also imperative to address the environmental impact of the ICT sector in terms of carbon emissions. If the growth rate of carbon emissions from the ICT sector matches that of the period between 2007 and 2020, global greenhouse gas emissions associated with this sector will increase from 3.0–3.6% in 2020 to 14% of all emissions by 2040 [58]. This projection emphasizes the urgent need for strategies to mitigate the sector's carbon footprint. The rising cost of energy and the global push for low carbon emission policies further intensify the pressure on companies and individuals to adapt and prioritize energy efficiency measures.

To address some of these challenges, more energy-efficient, single-ISA Asymmetric Multicore Processors (AMP) have been developed, such as ARM's big.LITTLE (2011), Apple's M-family of processors(2020), and Intel's Alder Lake (2021) and Raptor Lake (2022) family of processors. These processors combine high-performance cores with large caches (so-called *p-cores*) and small energy-efficient cores (*e-cores*) to deliver both high performance and energy efficiency. This work centers around Intel's Alder Lake processors, specifically taking into account the popularity of their CPUs, which accounts for almost 63 % of market share worldwide [17].

While efficient scheduling of threads on AMPs is critical, it is complicated. For example, Cao et al. [11] showed that naively replacing p-cores with e-cores has a negative effect on performance and energy efficiency. As a result, methods of efficiently scheduling threads on AMPs have attracted much attention in the literature (e.g., [39] and [54] and Section 7).

Java Virtual Machines (JVMs) spawn lots of worker threads to perform garbage collection (GC) which are completely out of the programmer's control, which presents an opportunity for automatic scheduling of these activities. This is especially alluring taking two factors into account: (1) GC is present by default in every running instance of a JVM and (2) pervasive adoption of JVMs in the cloud.i The primary responsibility of GC is to manage the allocation and deallocation of memory. Among all GCs available on the JVM, ZGC, and Shenandoah are the least energy-efficient due to the additional synchronization cost of running GC operations in parallel with the application [52].

In this work, we explore the energy impact of scheduling GC threads on e-cores. Specifically, we investigate the energy consumption of applications and evaluate the benefits of a fully concurrent GC, ZGC, running on a heterogeneous Alder Lake. The idea of scheduling GC on e-cores was explored in the past [11] (referred to subsequently as the YinYang study). This study concluded that in terms of energy efficiency, GC threads are best suited for small clocked-down in-order cores as they are bound to wait for memory most of the time (see also [38]). They observed a decrease in the energy for the entire application on average by 11 % without significant performance degradation when GC is executed on e-cores instead of p-cores. Since then, the growing demand for low-latency server applications led to the emergence of fully concurrent GCs. Furthermore, commodity AMP systems now boast more cores than previously anticipated by the YinYang study. In light of these advancements, our objective is to reassess previous conclusions by considering the availability of high-core server AMP machines and a concurrent GC, with a specific emphasis on latency constraints. Additionally, we will delve into memory-related aspects associated with transitioning from p-cores to e-cores, an area that has been overlooked in prior studies.

Our initial hypothesis is that scheduling a fully concurrent GC on e-cores can result in energy reduction for applications without impacting performance or latency. To test this hypothesis, we investigated the following research questions:

**RQ1** What is the impact on the energy of executing ZGC worker threads on e-cores instead of p-cores on commodity x86 hardware?

**RQ2** What are the effects on performance?

**RQ3** What are the trade-offs, if any, between energy reduction, performance parameters, and memory used by applications?

The rest of the paper is organized in the following manner: we cover the most related work and other related concepts in Section 2; Section 3 presents the main idea; Section 4 describes our methodology, Section 5 shows our results; Section 6 explains the possible causes of differences in the results between this research and YinYang and Section 7 surveys additional related work.

## 2 Background

The background section of this paper delves into the foundational research that forms the basis for our current study. Specifically, we focus on the GC algorithm used in the prior research, namely Concurrent Mark-Sweep (CMS). We will also briefly explain the fundamental concepts, characteristics, and idiosyncrasies of the GC we will employ in this study, namely, ZGC. Furthermore, this section provides essential information about Alder Lake, the platform we use in our experiments, to establish a context for our research.

### 2.1 Previous Work on VM Activity Scheduling and Energy Efficiency

Our work builds on the findings of YinYang [11]. This work evaluated the energy efficiency of scheduling various VM activities, including GC, just-in-time (JIT) compilation, interpreter, and application threads, on p-cores and e-cores.

YinYang's approach was evaluated using the CMS GC (see Section 2.2) using an AMP system comprising a single 2.8 GHz Phenom II core (p-core) and two 1.6 GHz AtomD cores (e-cores) which were connected through PCI Parallel Port Card and Arduino board. The main reason for this "constructor" design was the lack of readily available AMPs on the market at the time.

In addition, the YinYang study explored future systems in which adding more big cores is not feasible because of power or energy constraints. Thus, the created system contained only 1 p-core and 2 e-cores. However, with the advent of "off-the-shelf"

AMP designs, which incorporate multiple high-performance cores alongside lower-performance energy-efficient cores, adding more big cores is no longer infeasible as was anticipated.

## 2.2 Mostly-Concurrent GC (CMS) vs "Fully" Concurrent GC (ZGC)

In this section, we delve into the essential differentiation between a mostly-concurrent GC, exemplified by CMS, and a fully concurrent[1] GC, represented by ZGC.

Currently, there are two fully concurrent GCs in OpenJDK: ZGC and Shenandoah. We leave Shenandoah [15] for future work and instead focus on ZGC. However, the main principles discussed in this paper should hold true for Shenandoah as well. By addressing the distinction between mostly-concurrent and fully concurrent collectors, we aim to underscore the significance of adopting a GC like ZGC in our research, which allows us to add a new dimension, namely the latency aspect, to the problem of energy-efficient scheduling of GC on AMPs. Through this exploration, we establish the groundwork for comprehending the pivotal role of fully concurrent GCs in the context of AMPs.

### 2.2.1 Concurrent Mark and Sweep (CMS)

CMS performs most but not all GC work concurrently with the program. To coordinate work, CMS uses brief stop-the-world (STW) pauses that can become longer, for example when collection in the young generation happens while collection in the old generation collection is already ongoing. To avoid long pauses, CMS does not move objects during reclamation, which makes it sensitive to fragmentation. Due to these limitations (being fragmentation prone as well as rear but unpredictable long STW pauses), CMS was deprecated in Java 9 and removed in Java 14.

### 2.2.2 The Z Garbage Collector (ZGC)

ZGC is a concurrent collector available in OpenJDK first as a single-generation collector, and since OpenJDK 21, as a generational collector.

**ZGC Basics**    ZGC [33] is a low-latency, parallel, concurrent, compacting, generational GC. It implements algorithms whose STW pause times do not increase with the size of the heap, including concurrent evacuation of pages during reclamation (meaning regions of memory are freed by moving all live objects away from the page). Its high-level algorithm was described by Yang and Wrigstad [61] (for single generation). Generational ZGC has two types of cycles: minor — collecting young generations only, and major — collecting young and old generations at the same time. Minor and major collections can run in parallel.

To enable concurrent compaction, ZGC uses load barriers to trap accesses to relocated objects and remap dangling pointers to point to their updated location before accesses may commence. The ZGC algorithm forces all pointers on the heap to be remapped once per GC cycle and uses atomic operations such as compare and set to coordinate GC workers and program threads operating concurrently on the same objects. Thus, ZGC has an intricate interaction with mutators.

---

[1] Technically, ZGC is not fully concurrent as it stops the program several times during a GC cycle, but none of those pauses perform work that is proportional to the size of the heap.

**Impact of ZGC on Application Performance**  As long as the concurrent worker threads can keep up with a program's allocation rate, GC in ZGC activities never causes mutators to block (modulo brief STW pauses where no real GC work takes place). The need for constant coordination between mutators and GC however introduces additional checks and mutators occasionally perform object relocation instead of waiting for the GC to do so. Every phase change (e.g., from marking to relocation) forces mutators to check that all pointers are still valid (on the first subsequent access), causing a wave of mutators to hit a slow-path of load barriers, and potentially causing a slowdown in application execution time (throughput).

For example, [60] and [1] showed that reducing GC frequency can negatively impact data spatial locality. This decrease in data spatial locality can lead to a decline in overall execution time. Hence, there exists a delicate balance between the work performed by a GC and the execution time of an application.

One crucial consequence of performing all GC activities concurrent with the application, compared to the mostly-concurrent GC's, is that slowing down GC worker threads should not increase application latency given enough headroom (at deployment), as mutators never block on GC operations.[2]

**ZGC Heuristics**  The goal of a concurrent collector is for the reclamation rate to match the allocation rate of the application while minimizing the impact on performance. To that end, ZGC uses non-trivial heuristics to determine when to start a GC cycle to prevent Out-Of-Memory (OOM) errors, how many threads should be utilized for each cycle, etc.

There are various triggers for GC cycles, including high allocation rate, high heap usage, or a lack of collection activity for a certain period, such as 5 minutes. Additionally, ZGC may occasionally collect the old generation even without specific triggers. These heuristics take into account the available free memory, the projected time until an OOM error occurs based on average allocation rates, and accounts for unforeseen changes.

To determine the appropriate number of GC workers needed to prevent OOM errors, ZGC analyzes the durations of previous GC cycles and adjusts the worker count based on available hardware. Furthermore, ZGC predicts the duration of the next GC cycle based on the number of GC workers and calculates the optimal start time for that cycle.

---

[2] By enough headroom we mean that there should be enough available CPU and RAM. Otherwise, a concurrent collector risks constant stalling, as it is not able to react to the allocation behavior of the application.

### 2.3 Intel Alder Lake

At the beginning of this study (spring 2022), AlderLake was the only x86 AMP chip on the market.

Alder Lake is the code name for the 12th generation of Intel's core line of processors. It consists of p-cores (Golden Cove) and e-cores (Gracemont). The Golden Cove cores support Hyper-Threading, allowing two threads to run on a single core, while the Gracemont cores are single-threaded with a smaller out-of-order window than on p-cores (Figure 1). Gracemont cores are organized in clusters or modules, each comprising four e-cores. The space occupied by one such module is comparable to that of a single p-core. It's noteworthy that, due to resource-sharing among the four e-cores within a module, even if only one e-core is active, the entire module must be powered up.

This architecture supports Intel's Thread Director technology based on the Enhanced Hardware Feedback Interface (HFI). This hardware-based technology provides enhanced telemetry data about the state of the core to the Operating System (OS). It uses a shared table between the hardware and the operating system. The table's contents may be updated due to changes in the operating conditions of the system (e.g., reaching a thermal limit) or the action of external factors (e.g., changes in the thermal design power). The information that HFI provides is numeric, unitless capabilities relative to other CPUs in the system. These capabilities have a range of [0,255] where higher numbers represent higher capabilities. Energy efficiency and performance are reported in separate capabilities. If either the performance or energy capabilities of a CPU are 0, the hardware recommends not scheduling any tasks on such CPU for performance, energy efficiency, or thermal reasons, respectively.

The Thread Director can detect the instruction mix (scalar/vector) used in any given thread. Typically, vector/AI workloads will be prioritized to performance cores while scalar instructions and background tasks (e.g., being in the background as opposed to the foreground in gaming) are moved to efficiency cores.

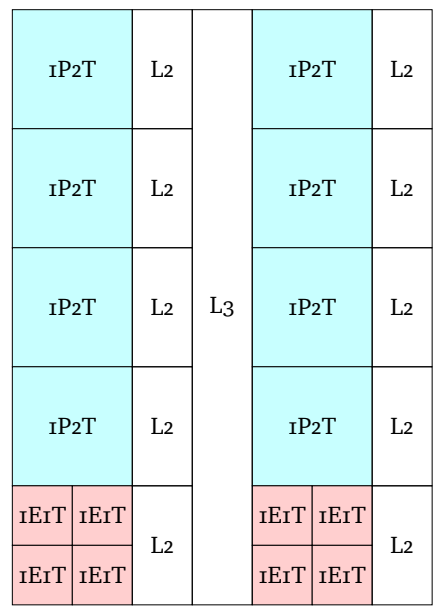

■ **Figure 1** Schematic architecture of Alder Lake. 1P2T stands for one p-core (PC) and two hardware threads (T). 1E1T: one e-core and 1 hardware thread. LLC: last-level cache.

The Linux kernel supports the Thread Director since version 5.18 [23, 24]. The x86_energy_perf_policy utility is a tool for managing energy-performance policy settings on Intel Architecture Processors. These settings can be controlled via Model Specific Register (MSR) updates, allowing users to influence how aggressively the system saves energy. The utility has been available since version 5.19.

The main policy setting, MSR_IA32_ENERGY_PERF_BIAS (EPB), plays a crucial role in hardware decisions related to CPU idle states (C-states) and Processor Performance States (P-states). By providing a hint to the hardware, EPB determines the level of aggressiveness with which the system implements the C-state and P-state selections made by the operating system. This, in turn, influences the trade-off between energy efficiency and performance. The EPB range spans from 0 to 15, where 0 represents maximum performance mode without sacrificing any performance for energy efficiency, while 15 indicates a policy that allows a measurable performance degradation to maximize energy efficiency.

## 3   Key Objective: Evaluating ZGC Scheduling on Energy-Efficient Cores

The primary objective of this work is to extend the findings of the YinYang [11] study in a contemporary context, using ZGC and commodity hardware. Specifically, we aim to investigate the impact of scheduling GC workers on e-cores versus on p-cores cores in terms of energy, performance, latency, and memory usage.

Inspired by the YinYang results [11], we anticipate the following:

**Hypothesis 1:** *Given that GC work in fully concurrent collectors isn't on the critical path, intentionally scheduling it on slower e-cores should not have a detrimental impact on application performance or latency. Instead, this approach should lead to energy savings.*

Our experimental setup aims to simulate real-world scenarios. Thus, we have specifically selected workloads with CPU utilization levels below 50%, as it is common practice to deploy latency-oriented benchmarks while keeping machines underutilized [5], e.g., to be able to handle sudden spikes without breaking SLAs, etc. By adopting this approach, we ensure that both mutators and GC can share system resources without contention. This plays a critical role, as it enables us to allocate GC threads separately to different types of cores without adversely impacting mutators' performance.

To facilitate comparison between different scenarios of GC scheduling, we pin each thread to a specific core type upon its creation using the *sched_setaffinity* kernel call [34]. We do this from inside a JVM. Once pinned, we do not repin threads until the program has finished running. To enhance our ability to accurately attribute measurement changes to GC, we have chosen to pin all threads, including ZGC threads, application threads, and VM threads. By doing so, we gain greater control over the experimental setup. For instance, if we allowed mutators to move between p-core and e-cores, it would become challenging to definitively attribute any observed energy reduction solely to GC. Therefore, the decision to pin all threads enables us to isolate and measure the impact of GC more precisely, facilitating a more reliable analysis of energy efficiency.

■ **Table 1** Parings of hardware configurations to be compared. For each hardware configuration, the table shows the number of hardware threads (HWTs) and cores as well as the L2 cache size. The total number of cores is chosen to use the maximum possible number of e-cores, up to available 8. For example, to test 1:1 HWTs ratio, we can compare placing GC on 4 p-cores with 8 HWTs vs 8 e-cores with 8 HWTs: 8:8 → 1:1. We wanted to test bigger configurations as the strength of e-cores in quantity.

How to read hardware configuration names: The names of each configuration consist of 2 symbols, for example, 4E. Numbers represent the number of cores used for GC. The letters represent a type of core. E stands for e-cores, and P is for p-cores. So 4E means that GC runs on 4 e-cores. Note, that other threads always execute on 4 p-cores. When we compare 2 configurations we separate them with a slash. We use the configurations after the slash for normalization.

| Configuration 1 | | | Vs. | Configuration 2 | | | HWTs ratio | Name |
|---|---|---|---|---|---|---|---|---|
| HWTs | Cores | L2 Cache MB | | HWTs | Cores | L2 Cache MB | | |
| 4 | 2P | 2.5 | | 8 | 4P | 5 | 1:2 | 2P/4P |
| | | | | 6 | 6E | 4 | 2:3 | 6E/2P |
| | | | | 8 | 8E | 4 | 1:2 | 8E/2P |
| 8 | 4P | 5 | | 4 | 4E | 2 | 2:1 | 4E/4P |
| | | | | 6 | 6E | 4 | 4:3 | 6E/4P |
| | | | | 8 | 8E | 4 | 1:1 | 8E/4P |

## 4 Methodology

In this section, we will provide a comprehensive overview of the experimental setup and methodology employed in our study. We will delve into the hardware and software aspects, detailing the hardware differences between hardware configurations used for our experiments. Additionally, we will explain our benchmark methodology, outlining the specific benchmarks chosen and the reasons behind their selection. We will also address the crucial aspect of heap sizing and its significance in our research. To ensure stable and reliable results, we will discuss our approach involving iterations and cache flushing. Furthermore, we will explain which measurements we collected, including the reasons for measuring specific metrics and the methods employed to obtain them. Lastly, we will explore the statistical analysis techniques utilized and justify their suitability for our study. This section serves as a fundamental foundation for the subsequent discussions and findings presented in the paper.

### 4.1 Hardware and Software

The Alder Lake system we use for our experiments has a 12th Gen Intel Core i9-12900K processor. The i9 chip features a total of 16 cores, comprising a combination of 8 p-cores (performance cores) and 8 e-cores (efficiency cores). Each p-core has two hardware threads (hyper threads), in contrast with e-cores which have one hardware thread each. Each p-core has 80 KB of L1 cache and 1.25 MB of L2 cache. E-cores have 96 KB L1 cache and share a 2 MB L2 cache per 4 e-core module (Figure 1). The processor incorporates a shared L3 cache of 30 MB. It is shared between both p-cores and e-cores. The system has 128 GB of RAM. Both p- and e-cores can run with the same frequency up to 8500 MHz.

We ran our experiments on Ubuntu 22.04 with Linux 5.19 kernel. We evaluated both EPB set to 15 (power save mode) and EPB set to 0 (performance mode). As anticipated, power save mode reduces the energy consumption of applications, primarily affecting threads running on e-cores. Moreover, we did not observe reduced execution time when running an application in performance mode on p-cores vs. power save mode on e-cores, only increased energy differences. As our goal is energy reduction, we decided to focus our comparison with EPB 15 (power save mode).

We also tested the Linux 5.18 kernel but chose 5.19 since it gives more capabilities to a user to control the environment to target energy efficiency.

All the measurements were performed with OpenJDK 20 and generational ZGC as a baseline [42].

### 4.1.1 Hardware Configurations

To gain insight into how AMPs can be utilized for energy savings of Java programs, we organized our investigation by comparing configurations listed in Table 1 based on the relation of hardware threads. For example, a 1:1 ratio results in the 4P-8E hardware configuration since each p-core has two hardware threads. We chose to investigate various ratios to determine the optimal balance between the use of e-cores and p-cores.

Note, that we only vary how GC threads are scheduled, either on p or e-cores, which is reflected in the hardware configuration names. All other threads are always scheduled on 4 p-cores.

Table 1 lists different L2 sizes for each hardware configuration. Since e- and p-cores do not share L2 caches, in the case of using 4 p-cores (4P), GC gets 5 MB L2, and 4 MB in the case of 8 e-cores (8E). (Note that as GC runs on dedicated cores, it does not need to compete with other program threads for L1 and L2 cache.)

An important point to consider is that the JVM lacks awareness of the underlying hardware architecture and the exact count of available processor cores. Consequently, it may allocate more or fewer GC worker threads than the actual core count. For instance, in a 1P configuration, ZGC tends to utilize more GC worker threads on average than what a single p-core with 2 hardware threads can effectively handle. To maintain the accuracy of our analysis, we decided to exclude this specific configuration from our study. This choice was made because it becomes challenging to estimate the energy impact accurately when GC worker threads compete for CPU resources.

## 4.2 Benchmark Methodology

We selected specific benchmarks (BMs) that provide insights into latency, as ZGC's goal is to minimize (tail) latency, and not throughput. The chosen benchmarks are Hazelcast [19], and a recent, pre-release version of the DaCapo suite [10].

■ **Table 2** Benchmark Parameters. #: iterations per run. We repeat each run ten times for stability. Avg CU: Average CPU utilization. Threads: number of threads used by the application. * denotes default, which means using the entire machine, except for bioJava which is single-threaded.

| BMs | Size | Threads | Full Name | # | ∼Avg CU(%) |
|-----|------|---------|-----------|---|------------|
| tomcat | default | 2 | tomcat_def_t2 | 20 | 12 |
|  | default | 4 | tomcat_def_t4 | 20 | 23 |
|  | large | 2 | tomcat_large_t2 | 10 | 12 |
|  | large | 4 | tomcat_large_t4 | 10 | 24 |
| lusearch | default | 2 | lusearch_def_t2 | 20 | 12 |
|  | default | 4 | lusearch_def_t4 | 20 | 19 |
|  | large | 2 | lusearch_large_t2 | 10 | 11 |
|  | large | 4 | lusearch_large_t4 | 10 | 19 |
| spring | default | 2 | spring_def_t2 | 20 | 12 |
|  | large | 2 | spring_large_t2 | 10 | 17 |
| luindex | default | * | luindex_def | 20 | 6 |
|  | large | * | luindex_large | 10 | 8 |
| fop | default | * | fop_def | 20 | 16 |
| bioJava | default | * | bioJava_def | 20 | 9 |
|  | large | * | bioJava_large | 10 | 10 |
| hazelcast | 100000 keys | * | hazelcast_100 | 1 | 24 |
|  | 20000 keys | * | hazelcast_20 | 1 | 14 |

### 4.2.1 Benchmark Descriptions

**Hazelcast: Real-Time Stream Processing**  Hazelcast is a framework designed for real-time stream processing. To ensure consistent and meaningful comparisons, we followed the parameters suggested by Topolnik [55]. The workload of Hazelcast is determined by the size of its key set. To explore the impact of different workloads, we conducted experiments using key-set sizes of 100000 and 20000 (the former is the higher workload).

**DaCapo: Tomcat, Lusearch, Spring, Luindex, Fop, and BioJava**  When we use the term DaCapo in this paper, we refer to the Chopin development branch (ee242f22). We run the latency-oriented benchmarks within the suite: Tomcat, Lusearch, and Spring. We excluded Kafka and JME due to a low CPU utilization issue (noted by the benchmark maintainers[3]). We also excluded H2 due to a memory leak resulting in longer execution times that we were able to reproduce across multiple machines and garbage collectors. In addition, we did not run Tomcat, Spring, and Lusearch with small input sizes, as the GC activity in these benchmarks was very low. We also explored throughput-oriented benchmarks with a high GC level: Luindex, Fop, and BioJava. To explore various CPU utilization rates, we modify the data set size and the number of application threads (-t parameter) whenever applicable. For example, some of the mentioned benchmarks are not inherently scalable, meaning their CPU utilization does not vary with an increase in the number of mutators or input size. Hence, we selected multiple thread configurations of the same benchmark only if they exhibited diverse CPU behavior. For more detailed information on the benchmarks, we refer to Table 2.

---

[3] https://github.com/dacapobench/dacapobench/blob/dev-chopin/benchmarks/status.md. Last accessed 2024-02-08.

### 4.2.2 Heap Sizing

We carefully selected the heap size for each benchmark to avoid allocation or relocation stalls as well as OOM errors, which should not happen under proper deployment, and that would severely complicate analysing the results as the behaviour of ZGC changes when responding to a stall. The elimination of stalls is of utmost importance for workloads that require low latency, as it ensures that the program never blocks on the GC. Larger heap sizes were not extensively explored as they resulted in less GC activity.

Starting from a base size of 16 MB, we executed each application with varying heap sizes. If a particular heap size resulted in allocation or relocation stalls, or OOM errors, we increased the heap size by 10%. We picked the smallest heap size that allowed the successful completion of a benchmark in five consecutive runs, i.e., without stalls or memory failures.

Note that each configuration in our study has a unique heap size. This distinction arises from two primary reasons. First, we aimed to investigate the variations in memory requirements when deploying applications on p-cores compared to e-cores. The variations mentioned are bound to occur, resulting from differences in cache sizes, core frequencies, and the utilization of hyperthreading.

Second, we sought to establish a fair comparison among vastly different configurations. Setting the heap size to the largest value across all configurations would result in some configurations experiencing no GC activity at all. It is practically impossible to find a single heap size that yields identical GC activity across all configurations. Therefore, we identified the minimum possible heap sizes where mutators never block on GC and reported the corresponding GC activity for the comparisons (Table 2).

### 4.3 Measurements

While our primary focus in this work is on energy consumption, we also investigate the relationship between energy and other performance parameters, namely throughput and latency (**RQ1**, **RQ2**). While energy efficiency is crucial for assessing the environmental impact and overall sustainability, evaluating throughput and latency is equally vital in understanding the system's responsiveness and user experience. It is important to consider how energy optimizations may affect performance and latency, as solutions with significant drops in these areas may not be as practical or attractive for real-world applications. Striking a balance between energy efficiency and maintaining acceptable performance and latency levels is crucial for the practical deployment of energy optimization techniques.

### 4.3.1 Measuring Latency

The pause times of ZGC are not a reliable approximation for overall latency as the work carried out during these pauses is minimal, and not proportional to the the application's memory. Instead, to accurately capture latency and response times, we employed benchmarks (described in Section 4.2) that provide application-specific latency measurements.

The DaCapo benchmarks provide two types of latency measurements: simple latency and metered latency [64]. For our analysis, we specifically concentrate on the 99.9th percentile metered latency, which closely represents the real response time experienced by the application, taking into account factors beyond GC pauses. To ensure consistency, we also report the 99.9th percentile latency for Hazelcast, aligning with the approach taken in the DaCapo benchmarks.

### 4.3.2 Measuring Throughput

In addition to latency, we also measure and report throughput. While ZGC aims to deliver low tail latency and not high throughput, the latter still needs to be within acceptable boundaries. In this study, we measure throughput by evaluating the execution time of the benchmarks.

### 4.3.3 CPU Utilization

We used vmstat [21] to monitor CPU utilization. We consider CPU utilization as an important factor in our evaluation. Although it is not explicitly presented in the results, CPU utilization plays a crucial role in determining the available computational resources that the GC can utilize without application interference.

As already mentioned, in order to meet SLA guarantees, deploying latency-sensitive workloads requires a certain level of CPU headroom in the system. However, there are no established guidelines on what constitutes an ideal amount of headroom except that *proper deployment* means that a machine is underutilized allowing ZGC to execute without interfering with mutators. Therefore, we conducted experiments with different CPU utilization rates to cover a range of use cases and gain insights into the impact of CPU utilization on the performance of the Java applications under investigation.

### 4.3.4 Measuring Energy Consumption

We measure energy consumption using RAPL (Running Average Power Limit) [22], a feature supported by recent Intel systems. RAPL provides an interface to read machine-specific registers (MSRs) that estimate the energy consumption across various domains of the system.

RAPL categorizes energy consumption into four parts: *PKG* represents the entire package, which includes cores, shared caches, memory controllers, and optional uncore devices; *PP0* refers to the cores themselves; *PP1* corresponds to the uncore devices, typically including components such as GPUs; and *DRAM* represents the memory subsystem.

It is important to note that the Alder Lake machine used in our experiments only supports the measurement of *PPo* and *PKG* energy consumption. Therefore, we utilized these measurements to report the energy scores in our study. The methodology and technique for measuring energy consumption were similar to the approach outlined by Shimchenko et al. [52].

### 4.3.5 GC-Related Metrics

Throughout the paper, we mention three different GC metrics: (1) cycles, (2) time, and (3) GC activity. Cycles refer to a number of the total number of cycles (minor and major). Time refers to the wall-clock time of executing mark and relocate phases, which we use as an approximation of the total time spent doing GC because these two phases dominate a ZGC cycle. Activity is wall-clock time spent doing GC divided by the total wall-clock time of an application. We record all three metrics to understand which parameters correlate with energy to be able to explain energy variances better.

### 4.3.6 Ensuring Stable Results: Iterations and Cache Flushing

To achieve stable and reliable results, we employed several techniques. First, we specified the number of iterations per run for each benchmark (indicated in the 4th column of Table 2) to achieve a steady state [9]. We divided iterations into a warm-up phase consisting of several iterations, followed by a measurement phase. The warm-up phase allowed the benchmark to reach a steady state before capturing the performance metrics. The last five iterations were considered the measurement phase, during which we collected data for analysis. We considered a state to be steady when a coefficient of variance in regard to execution time is below 0.05. In the case of Hazelcast, which is a long-running benchmark, we assumed that any initial jitter or variability would be smoothed out over the duration of the benchmark. Therefore, we focused our analysis on the overall performance trends and measurements collected throughout the entire run.

Between each iteration, a cache flush was executed to reset the states of the L1, L2, and L3 caches. This step ensures that every iteration commences with a pristine cache state, eliminating any lingering data from preceding iterations that might otherwise impact performance. By doing so, each iteration initiates with a clean slate, akin to being executed independently. The primary objective is to eliminate noise attributed to the VM while maintaining the integrity of each application iteration, treating it as if it ran autonomously.

To further enhance the reliability of our measurements we started each benchmark a minimum of 10 times with different JVM instances, with multiple iterations per run. This counters variance that occurs across different JVM instances, which can have a considerable impact on performance. Please refer to Table 6 in Appendix A to see the relative standard deviation (RSD) for energy data, which shows that our approach led to RSD mean to be 1.21 %.

Ultimately, we refrained from triggering explicit GC between benchmark iterations to avoid disrupting ZGC heuristics. This decision does not contradict our choice to

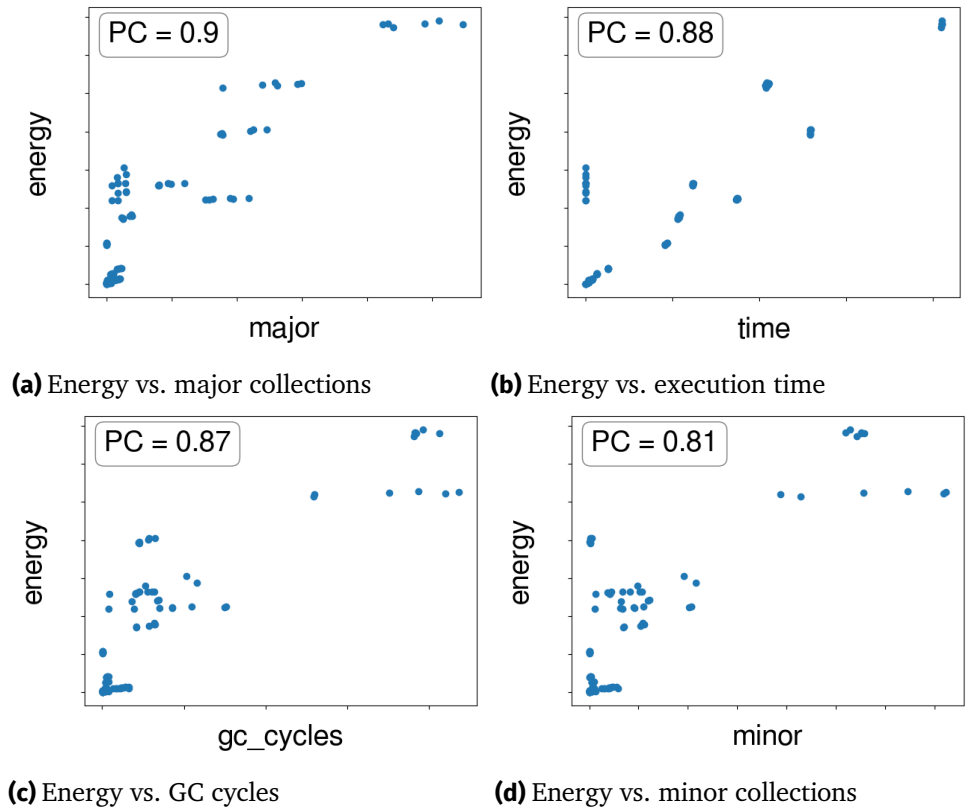

**(a)** Energy vs. major collections

**(b)** Energy vs. execution time

**(c)** Energy vs. GC cycles

**(d)** Energy vs. minor collections

■ **Figure 2** PC correlation above 0.8 for energy. The graph shows that energy has a high correlation with the number of major and minor collections, execution time, and the number of GC cycles, which is the sum of minor and major collections.

execute cache flushes. Our rationale is rooted in the nature of ZGC, which excels in comprehending patterns within long-running server applications—its typical deployment scenario. In contrast, our benchmarks operate in a short-lived context, an environment atypical for ZGC. By abstaining from explicit GC calls and allowing ZGC heuristics to function undisturbed, especially during the final stable iterations, we strive to emulate a more authentic ZGC runtime experience.

### 4.4 Statistical Analysis

We performed a statistical analysis of the benchmark results using Welch's t-test [59], Grubb's outlier test [20], and Yuen's t-test [63]. Welch's and Yuen's t-tests were used because we cannot assume a normal distribution of data or equal variances. Grubb's test identified outliers, and if present, Yuen's test was used. A significance level of 0.05 was used, and p-values determined whether the data sets exhibited significant differences.

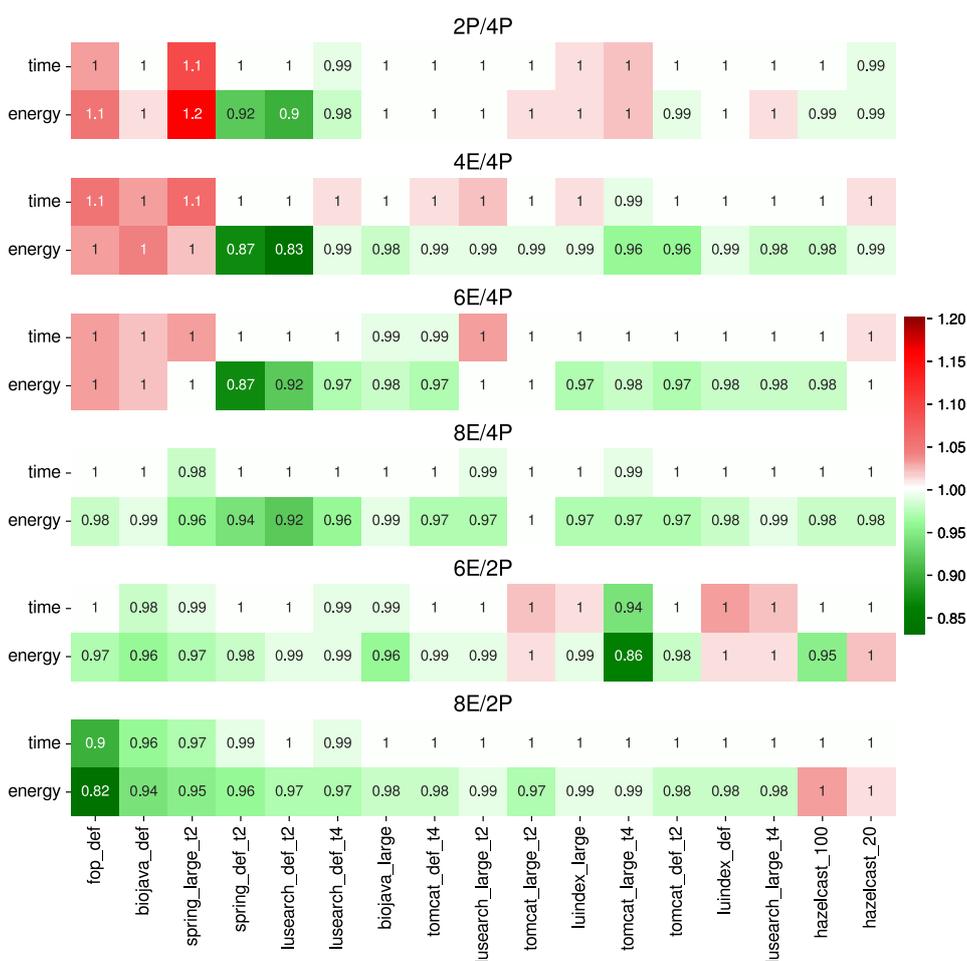

■ **Figure 3** Normalized execution time (time) and energy consumption. Use Table 1 to consult how to interpret comparison names: 2P/4P, 4E/4P, 6E/4P, 8E/4P, 6E/2P, 8E/2P. The BMs are sorted by energy for 2P configuration. The color scheme: green-red ⟶ [0.5;2].

**Main conclusions from the heatmap:** the e-configurations demonstrate higher energy efficiency than p-configurations while not significantly impacting overall execution time. In addition, the 8E/2P (2:1 hardware thread ratio) comparison stands out as the most energy-efficient ratio.

## 5 Results

In this section, we present an analysis of the key performance metrics measured during our experiments: energy consumption, execution time, latency, and memory utilization.

■ **Table 3** Statistical analysis of energy and throughput data (refer to Section 4.4 for more details on methodology.)

**Main conclusion – energy:** while we cannot assert any statistically significant differences in energy consumption between the 2P and 4P configurations as well ass 4E and 4P, it is worth noting that other configurations do meet the 0.05 significance level. Moreover, 4E/4P configuration has a positive confidence interval range, which implies that there might be a positive mean difference between the two groups.

**Main conclusion – throughput:** as the calculated p-values consistently exceed the 0.05 significance level except of 4E/4P, our analysis indicates mostly a lack of statistical evidence supporting performance differences between GC running on p vs. e-cores.

| Energy | | | |
|---|---|---|---|
| Comparison | P-value | Confidence interval | Improvement |
| 2P/4P | 0.985 | (−0.024, 0.023) | 0.0005±0.0235 |
| 4E/4P | 0.072 | (0.003, 0.057) | 0.03±0.027 |
| 6E/4P | 0.028 | (0.007 0.044) | 0.025±0.0185 |
| 8E/4P | 0.000072 | (0.02, 0.04) | 0.030±0.010 |
| 6E/2P | 0.016 | (0.01, 0.045) | 0.0275±0.0175 |
| 8E/2P | 0.027 | (0.009, 0.056) | 0.0325±0.0235 |
| Execution Time | | | |
| 2P/4P | 0.129 | (−0.019, 0.001) | −0.009±0.0095 |
| 4E/4P | 0.028 | (−0.019, −0.003) | −0.011±0.008 |
| 6E/4P | 0.094 | (−0.012, −0.0001) | −0.006±0.006 |
| 8E/4P | 0.089 | (0.00009, 0.005) | 0.0035±0.0025 |
| 6E/2P | 0.470 | (−0.007, 0.011) | 0.002±0.009 |
| 8E/2P | 0.112 | (−0.001, 0.025) | 0.012±0.013 |

## 5.1 Energy

This section presents an analysis of the impact of different system configurations on energy consumption. By examining various comparisons of system configurations, we aim to understand how different setups affect energy consumption.

In this section, we address **RQ1**, which explores the impact on energy consumption of executing ZGC on e-cores versus p-cores. The aggregated energy results are presented in the second column of Figure 3, and demonstrate that e-configurations are generally more energy-efficient than p-configurations. To determine the parameter that most significantly influences energy consumption, we conducted a series of correlation tests, employing Pearson correlation (PC) [8] to identify linear dependencies.

Among all the parameters, energy exhibits the highest correlation with the number of major collections, while also being influenced by minor collections and the total number of GC cycles (as illustrated in Figure 2). Energy consumption of applications using ZGC has a high correlation with execution time, which supports findings from the previous study by Shimchenko et al. [52]. The duration of GC cycles and GC activity do not demonstrate a strong correlation with energy consumption, probably because both metrics are based on wall-clock time and not CPU time.

Figure 3 presents energy for six comparisons: 2P/4P, 4E/4P, 6E/4P, 8E/4P, 6E/2P, 8E/2P. Please refer to Table 1 on how to read the names of comparisons.

Notably, the e-configurations demonstrate higher energy efficiency than p-configurations, as illustrated in the second row. In the top heat map, the energy geomean for the 2P/4P (2P is normalized against 4P) configuration is 1, indicating that four fast hardware threads managed to keep up with the applications' allocation rate. We also did not observe statistically significant differences for this configuration, refer to Table 3. In the 6E/2P and 6E/4P configurations, the relative energy geomeans are 0.98, resulting in an overall energy reduction of $\approx 2\%$. Similarly, configurations 4E/4P, 8E/2P, and 8E/4P exhibit an overall energy reduction of about $\approx 3\%$. (The use of the geomean value follows the established practices for averaging normalized numbers [14].) For more precise information refer to Table 3.

One possible reason why the 6E configurations cannot achieve the same total energy reduction is that it consists of 1.5 e-core modules in terms of compute power (1 module contains 4 e-cores), but 2 modules need to be powered regardless. As a result, there is an energy penalty for powering more cores than can be fully utilized for computing.

We further analyzed how frequently each configuration achieved the maximum energy reduction. Table 4 in Appendix A summarizes for your convenience the maximum energy reduction of each benchmark from Figure 3. It shows that the 8E/2P (2:1 hardware thread ratio) configuration stands out as the most successful, showcasing the highest energy reduction in 10 out of 17 benchmarks. The 6E/2P configuration does not provide additional benefits compared to 8E/2P, similar to the comparison between 6E/4P and 4E/4P and 8E/4P. This reinforces the idea that employing half a module while powering the entire module does not yield observable advantages.

If we could identify the best hardware thread ratio between e and p-cores for each benchmark, we could achieve an overall energy reduction of $5.3\% \pm 0.0225$ or $\approx 5.5\%$, as indicated by the geomean value of maximum energy reduction from Table 4. As a potential future direction, leveraging machine learning to identify benchmark features that correlate most strongly with the choice of hardware thread ratios would be an interesting avenue to explore.

## 5.2 Latency

Figure 4 shows a latency distribution of latency-oriented benchmarks. It is important to highlight that the DaCapo benchmarks showcase high variability in their latency scores, which poses a challenge for establishing statistically significant findings. As a result, our approach was to assess the unaltered, non-normalized latency values instead of comparing specific configurations. By examining the non-normalized values, we can estimate the magnitude of the changes in latency scores, as the absolute value of latency is crucial in determining whether or not SLA constraints are met.

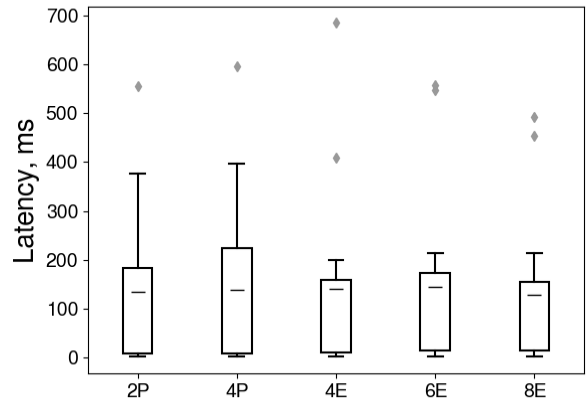

■ **Figure 4**  Latency (in ms) distribution for all benchmarks and tested configurations (note that latency values are not normalized; instead, they are presented for each configuration individually). The box contains values within the range between the 25th ($Q1$) to the 75th($Q2$) percentile. The horizontal line inside the box represents mean values. Dots are outliers. For a normal distribution, the Inter-Quartile Range ($IQR$) contains 50 % of the population, and $1.5 \times IQR$ contains about 99 %. It's an outlier if it is less than $Q1 - 1.5 \times IQR$ or if it is greater than $Q2 + 1.5 \times IQR$.

Based on our findings, the mean latency is not significantly affected by the transition from p to e-cores. The maximum latency, indicated by the end of the upper whiskers, remains effectively below the maximum latency observed in p-configurations. However, we do notice a distinction in outliers (aka tail-latency). Specifically, all latency scores of 4E are below 700 ms, while for every other configuration, it is below 600 ms, representing a 14 % reduction.

Nonetheless, considering our initial hypothesis 1, it appears that moving GC execution to e-cores does not substantially impact latency.

### 5.3 Throughput

The following two sections cover **RQ2**, i.e., the impact of executing ZGC on e-cores on performance and latency. The performance comparison presented in Figure 3 demonstrates that the scheduling GC threads on e-cores instead of p-cores does not significantly impact overall execution time, which is further supported by Table 3. This outcome aligns with our expectations, as when ZGC is properly deployed, it should not interfere with the critical path of mutators, allowing application threads to execute without disruptions. However, it is important to note that there may be natural variations in execution time due to the effects of GC on mutators, as discussed earlier.

For instance, we observed a statistically significant increase of 3 % in execution time for biojava_large in 4E/4P and 5 % for biojava_def in 4E/4P, as well as 9 % for fop_def in 2P/4P and 6 % in 4E/4P.

In the case of biojava_large in 4E/4P configuration, we observed 2.2× more GC cycles. More GC cycles lead to more phase changes where a phase is marking or relocating. Every phase change triggers a wave of mutators hitting slow paths in ZGC's load barriers, which can slow down the overall execution, as discussed in Section 2.

### 5.4 The Relationship between Hardware Resources, GC Workers, and Memory Requirements

Finally, we address **RQ3**, which involves investigating the memory tradeoffs associated with using e-cores instead of p-cores for ZGC. The memory usage for 2P, 4E, 6E, and 8E configurations, all normalized to 4P, is presented in Figure 5. The results clearly demonstrate that placing GC threads on e-cores requires more memory compared to using p-cores. For instance, deploying 4 e-cores for GC needs 8 % more memory than when using 4 p-cores.

To comprehend the reasons behind this memory difference, we analyzed the duration of a GC cycle (cycle time) and the number of GC workers (#workers) for each configuration (see Figure 5). Interestingly, the cycle time remains the same for e-configurations, but it is approximately 40 % higher than that of the 4P configuration. Meanwhile, the cycle duration in the 2P configuration is almost the same as that in the 4P configuration.

The fact that the number of GC workers decrease as the number of cores increase is due to a shortcoming in our prototype. ZGC's heuristics assume that all cores are available for doing GC (should the need arise) but we only permit GC work to be scheduled on select cores. ZGC's initial observation of faster cycles on configurations with more cores significantly influences its subsequent actions. When ZGC realizes that cycles are faster, it tends to reduce the number of GC workers more aggressively in configurations with more cores, assuming that fewer workers would be sufficient. On the other hand, in configurations with fewer cores, ZGC tries to use more GC workers to cope with the slower cycles.

It is reasonable to expect that having fewer GC workers could slow down the GC process, leading to increased memory usage on the 8E configuration compared to the 4E configuration, for example. In general, the 4E configuration executes 10 % fewer GC cycles than the 4P configuration, while the 8E configuration executes 20 % fewer GC cycles. When GC performs fewer cycles, yet each cycle maintains the same duration, it suggests that a smaller portion of memory is being cleaned up during the duration of an application. This, in turn, requires a larger amount of headroom in memory usage.

One plausible explanation for the notable slowdown of GC cycles on e-cores in comparison to p-cores is the substantial performance gap between the two. E-cores tend to operate at significantly lower speeds than p-cores and possess smaller reorder buffers, limiting the number of instructions that can take advantage of out-of-order execution. However, a comprehensive understanding of the factors driving this behavior requires additional investigation and analysis. It is also essential to explore potential avenues for fine-tuning GC heuristics to ensure that the number of GC workers scales proportionally with the available CPU resources.

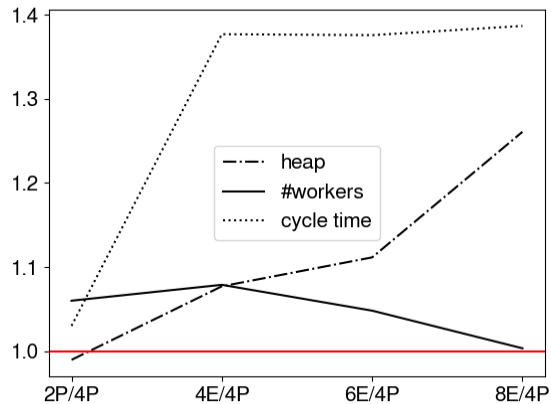

■ **Figure 5**  Geomeans of heap sizes, number of GC workers, and time per GC cycle for 2P, 4E, 6E, 8E normalized to 4P configuration.

## 6  Comparison with YinYang Study

In this section, we compare our energy results with the YinYang results [11]. Their experimental setup used an AMP system comprising a single 2.8 GHz Phenom II core (p-core) and two 1.66 GHz AtomD cores (e-cores). They compared this system with a single 2.8 GHz Phenom II core running mutators and another 2.8 GHz Phenom II core running GC. The Phenom II core does not support hyperthreading. Thus, in terms of hardware threads, they compared the 1:2 configuration.

In contrast to the 11 % observed in YinYang, we observe a 3 % energy efficiency improvement (in both cases without major performance loss) on a similar configuration.

Undoubtedly, comparing the numbers directly between our experiments and previous research presents many challenges. For example, we are using completely different GC algorithms. Also, our experiments involve p and e-cores from the same x86 family of processors. While there are architectural differences between Gracemont and Golden Clove, such as varying sizes of reorder buffers and caches, these differences are not as pronounced as those seen in the AMD vs. Intel combination studied by [11], especially taking into account a smaller speed/frequency difference (30 %) between cores in our experiments compared to YinYang (41 %).

Even though our results differed from previous research, reducing total energy by 3 % without requiring additional effort from developers represents a significant improvement.

## 7 Related Work

### 7.1 Core Sensitivity

One key difference between other work and this study is determining if a particular core is suitable for a certain workload/thread/process, namely, identifying core sensitivity. The basic approach is to consider a rate of performance (also known as a speed-up factor, SF) or energy improvement when executing a thread on one core type versus the other. There are two main approaches to finding out application core sensitivity: offline (static) or at run-time (dynamic). Offline approaches include running an application on all types of cores and then calculating an SF or energy metric to choose a Pareto-optimal solution [37, 48].

Run-time solutions may include hardware support. [44, 46] use a separate OS hardware-dependent module, which samples performance counters to estimate an SF on the fly. [56, 57] use a Performance Impact Estimation (PIE), a hardware-aided mechanism that has been shown to provide accurate SF estimates. On a software level, SF can be estimated by an IPC/IPS sampling method done on all core types periodically [6, 28, 29, 30, 56], combined with a history-based approach [12, 56] or an analytical model [50]. In addition to IPC, LLC misses proved useful for SF estimations [47, 51]. [43] also collects LLC misses with IPC and solves an Integer Linear Programming model to configure the scheduling for AMP.

Many studies combine offline profiling with dynamic run-time approaches. [62] and [49] profile available application to build a linear regression predicting model to estimate the rates at run-time by collecting performance counters. [18] and [35] predict SF based on offline collected statistics and online collected counters. The distance between an unknown thread and a profiled benchmark is measured as the difference in throughput on the core type the unknown thread is running on. [27] leverage compiler support to identify bottlenecks and a special hardware unit to schedule threads at a run-time.

Alternatively, instead of measuring an SF, one can leverage information about the nature of threads, as in this study. For example, [40] separate threads into OS-helper and application threads. The main assumption is that helper threads are not critical for performance and can run on power-efficient cores, while worker/application threads can benefit from running on big cores. Therefore, [40] suggests running OS threads on small cores while we experimented with GC threads.

### 7.2 Criticality

Another aspect of scheduling threads on asymmetric multi-cores is improving scalability for multi-threaded applications or, in other words, accelerating bottlenecks. The first step is to identify the bottleneck in sequential phases, which was explored by Saez et al. [45, 47]. In contrast, several other works [25, 26, 27, 53, 62] handle critical sections. There are multiple ways to accelerate a bottleneck, e.g., increasing core frequency [4], giving a thread higher priority in shared hardware resources [13, 26, 27], or migrating the bottleneck to a faster core with a more aggressive micro-architecture, or higher frequency [2, 25, 41, 62]. However, in the world of low-latency fully concurrent collectors, GC should not be critical by design or otherwise, deployment is incorrect. Therefore, we ignore this aspect in this work.

### 7.3 Optimization Metrics

The most important metric for optimization in this study is energy. In addition, many studies focused on optimizing throughput. [12, 50] prevent performance degradation as the system scales up by proposing an analytical model based on a sampling of hardware counters at a run-time. When a thread starts lagging, a scheduler will put it on a more efficient core. The objective of [6] is to avoid idle cores. [32] introduced an asymmetry-aware scheduler where the work assigned to each core is proportional to its processing power. Since ZGC is a low-latency GC, we did not focus on optimizing for throughput as it is not a intended use case for ZGC.

Another metric is fairness. Fairness guarantees equal progress for one or many applications. [31] defined it for all threads, while [32] extends the definition saying that only threads with the same priority should receive the same share of core processing power. Alternatively, the fair state can be defined based on the slowdown compared to an isolated run without any co-runner [56] – the slowdown should be the same for all the co-running applications. [48] extends the last definition to support DVFS as: *a scheduler is considered fair if the variation of performance degradation normalized to isolated run on a big core with the highest voltage and frequency is minimal*. This work complies with the second definition. As equal progress of mutators and GC threads is based on the fairness of the default Linux scheduler, optimizing for fairness is adjacent to this research. For example, [29] gives an equal share of big cores to all threads that benefit from it if their maximum performance degradation is less than a threshold. Otherwise, those threads receive extra time. [32, 62] adjust a time slice on big vs small cores. The slices of threads on big cores are relatively shorter than on little cores and threads are more often swapped on big cores. [27] ranks threads by their progress instruction count and gives a higher priority to the threads with lower progress to improve fairness.

#### 7.3.1 DVFS

Dynamic Voltage/Frequency Scaling (DVFS) is traditionally an orthogonal way of reducing energy compared to heterogeneous architectures. DFVS creates energy-efficient modes by lowering the voltage and frequency. However, many research works explored the combined efforts of AMP and DVFS. [11] investigated the frequency scaling on AMP for helper and worker threads in managed languages. They observed that the energy increases by more than 20% for GC and application threads with higher frequencies. They saw the potential to reduce energy even further by combining AMP and DVFS on small cores. [37] created an offline tool to schedule an application on AMP taking into account available levels of frequencies on each core. They concluded that AMPs are more efficient at saving energy than the traditional DVFS approach. However, combined together they deliver the most energy-efficient result. [50] is a load-balancing approach that is not limited to two types of cores. To achieve this, they proposed an analytical model, which considers architectural differences between available cores as well as frequencies. [48, 49] uses the DVFS approach not only to improve energy efficiency but also fairness by evening out the cluster's computing power. [36]' tool, VINCENT, implements a four-step approach that outperforms the built-in power management in Linux, achieving a remarkable 14.9% reduction in energy consumption. VINCENT's methodology involves identifying hot methods, profiling their energy consumption, evaluating different CPU frequencies, and scaling the top energy-consuming methods accordingly.

## 8 Conclusion

In conclusion, our study has provided insights into the energy efficiency of executing GC on e-cores versus p-cores. We have identified several key findings.

In our findings, we have initially demonstrated that transitioning GC execution from p-cores to e-cores can result in substantial energy savings, 5.3%±0.0225, provided that an appropriate hardware threads ratio is selected. Remarkably, without necessitating any additional developer effort, we observed a ≈ 3% reduction in energy consumption. This outcome underscores the practicality and real-world applicability of our approach. These energy efficiency gains contribute significantly to a more sustainable and efficient computing environment, emphasizing the value and relevance of our research findings.

Second, our investigation revealed that the transition to e-cores for GC does not result in any degradation in performance. Despite the shift in core utilization, the system maintained the same performance levels, ensuring that applications could operate at their intended speed and efficiency.

Our analysis of latency scores showed that the utilization of e-cores for GC can increase tail latency (e.g., by 15% running GC on 4 e-cores vs. 4 p-cores) which could impact the responsiveness of the system or violate SLA constraints. But overall e-cores manage to maintain the same latency as observed on p-cores.

Last, we found that the transition from p-cores to e-cores for GC requires more memory (e.g., by approx. 25% running GC on 8 e-cores vs. 4 p-cores), which is important if such a system is to be deployed in the real world.

All-in-all, our study contributes to the broader understanding of optimizing resource allocation in multi-core systems, paving the way for more sustainable and efficient computing practices.

**Acknowledgements** This work was funded by the Swedish Research Council grant 2020-05346 *Accelerated Managed Languages*, and the Swedish Foundation for Strategic Research SM19-0059 *Deploying Memory Management Research in the Mainstream*.

## A  Additional Tables

■ **Table 4**  Maximum energy reduction for each BM and a corresponding configuration, which achieved it. *total* indicates the number of BMs, for which a configuration showed the best results. By selecting the best result for each benchmark, we achieve energy reduction geomean of ≈ 5.5 %.

| Compared with | 4P | | | | 2P | |
| --- | --- | --- | --- | --- | --- | --- |
| BMs | 2P | 4E | 6E | 8E | 6E | 8E |
| hazelcast_20 | | 0.83 | | | | |
| hazelcast_100 | | 0.87 | 0.87 | | | |
| luindex_def | | | | 0.96 | | |
| lusearch_large_t4 | | | | 0.98 | | 0.98 |
| tomcat_def_t2 | | 0.96 | | | | |
| tomcat_large_t4 | | 0.98 | 0.98 | 0.98 | | |
| tomcat_def_t4 | | | 0.98 | 0.98 | 0.98 | 0.98 |
| lusearch_def_t4 | | | | 0.97 | | 0.97 |
| luindex_large | | 0.98 | 0.98 | | | |
| lusearch_def_t2 | | | 0.97 | 0.97 | | |
| lusearch_large_t2 | | 0.99 | | | 0.99 | 0.99 |
| spring_def_t2 | | | | | 0.96 | 0.96 |
| biojava_large | | | | | | 0.98 |
| tomcat_large_t2 | | | | | 0.97 | 0.97 |
| spring_large_t2 | | | | | | 0.95 |
| biojava_def | | | | | | 0.94 |
| fop_def | | | | | | 0.82 |
| total # BMs | 0 | 6 | 5 | 6 | 4 | 10 |

■ **Table 5** PC correlation between different parameters. PC is in the range [-1, 1], where 1 is a perfect positive correlation, -1 is a perfect negative correlation, and 0 is no correlation. Dark green cells highlight a high correlation [0.8, 1], and yellow cells highlight a moderate correlation [0.6, 0.8]. The matrix is symmetrical. The symbol "#" replaces the phrase "number of".

| Parameters | Energy | Exec. (t) | #GC | #Minor | #Major | Mark (t) | Relocate (t) | GC time | Heap per GC worker | Heap | GC workers | GC activity |
|---|---|---|---|---|---|---|---|---|---|---|---|---|
| Energy | 1.00 | 0.88 | 0.87 | 0.81 | 0.90 | 0.48 | 0.29 | −0.09 | −0.10 | −0.10 | −0.15 | −0.07 |
| Exec. (t) | 0.88 | 1.00 | 0.79 | 0.69 | 0.94 | 0.08 | −0.12 | −0.03 | 0.01 | −0.01 | −0.47 | −0.45 |
| #GC | 0.87 | 0.79 | 1.00 | 0.99 | 0.89 | 0.24 | 0.10 | −0.17 | −0.16 | −0.16 | −0.17 | −0.17 |
| #Minor | 0.81 | 0.69 | 0.99 | 1.00 | 0.81 | 0.26 | 0.14 | −0.16 | −0.15 | −0.15 | −0.11 | −0.11 |
| #Major | 0.90 | 0.94 | 0.89 | 0.81 | 1.00 | 0.15 | −0.02 | −0.18 | −0.17 | −0.17 | −0.32 | −0.33 |
| Mark (t) | 0.48 | 0.08 | 0.24 | 0.26 | 0.15 | 1.00 | 0.96 | 0.01 | −0.10 | −0.07 | 0.60 | 0.78 |
| Relocate (t) | 0.29 | −0.12 | 0.10 | 0.14 | −0.02 | 0.96 | 1.00 | 0.08 | −0.03 | 0.00 | 0.69 | 0.87 |
| GC (t) | −0.09 | −0.03 | −0.17 | −0.16 | −0.18 | 0.01 | 0.08 | 1.00 | 0.96 | 0.97 | −0.11 | −0.03 |
| Heap/GC worker | −0.10 | 0.01 | −0.16 | −0.15 | −0.17 | −0.10 | −0.03 | 0.96 | 1.00 | 1.00 | −0.18 | −0.14 |
| Heap | −0.10 | −0.01 | −0.16 | −0.15 | −0.17 | −0.07 | 0.00 | 0.97 | 1.00 | 1.00 | −0.16 | −0.11 |
| GC workers | −0.15 | −0.47 | −0.17 | −0.11 | −0.32 | 0.60 | 0.69 | −0.11 | −0.18 | −0.16 | 1.00 | 0.86 |
| GC activity | −0.07 | −0.45 | −0.17 | −0.11 | −0.33 | 0.78 | 0.87 | −0.03 | −0.14 | −0.11 | 0.86 | 1.00 |

■ **Table 6** The Relative Standard Deviation (RSD) measures the variability in collected energy data. Cases where RSD exceeds 5 % are highlighted in red. Notably, the benchmark 'fop' exhibits high variability, surpassing the 5 % threshold. This behavior is observed consistently across 80 iterations, indicating that 'fop,' being a short benchmark, does not achieve stabilization even with a substantial number of iterations.

| BM | 4P | 8E | 2P | 4E | 6E |
|---|---|---|---|---|---|
| fop_def | 4.79 | 3.64 | 5.91 | 6.36 | 3.04 |
| luindex_def | 1.37 | 0.33 | 0.94 | 0.73 | 0.52 |
| spring_large_t2 | 0.90 | 0.27 | 1.80 | 0.50 | 0.56 |
| hazelcast_100 | 0.52 | 0.36 | 0.70 | 0.52 | 0.25 |
| lusearch_large_t4 | 0.54 | 0.69 | 0.20 | 0.46 | 0.56 |
| luindex_large | 0.55 | 0.68 | 0.40 | 0.40 | 0.49 |
| tomcat_def_t4 | 0.30 | 0.38 | 0.78 | 0.53 | 0.79 |
| tomcat_large_t2 | 0.52 | 0.76 | 0.87 | 0.35 | 0.50 |
| lusearch_def_t2 | 0.91 | 1.06 | 1.03 | 1.19 | 0.51 |
| spring_def_t2 | 0.76 | 0.36 | 0.43 | 0.94 | 1.02 |
| tomcat_large_t4 | 1.19 | 0.61 | 0.89 | 0.86 | 0.23 |
| hazelcast_20 | 0.48 | 0.37 | 0.10 | 0.36 | 0.38 |
| biojava_def | 1.67 | 0.85 | 2.64 | 2.13 | 4.95 |
| lusearch_large_t2 | 1.46 | 1.41 | 0.74 | 1.41 | 0.80 |
| lusearch_def_t4 | 0.79 | 0.46 | 0.72 | 0.93 | 1.03 |
| tomcat_def_t2 | 0.85 | 0.91 | 0.32 | 0.69 | 1.18 |
| biojava_large | 1.65 | 2.03 | 3.21 | 1.74 | 2.52 |

## About the authors

**Marina Shimchenko** is a PhD student at Uppsala University. Contact her at marina.shimchenko@it.uu.se.
https://orcid.org/0000-0002-0701-8540

**Erik Österlund** is a Principal Engineer at Oracle in Sweden. Contact him at erik.osterlund@oracle.com.
https://orcid.org/0000-0003-3686-8568

**Tobias Wrigstad** is a professor in computing science at Uppsala University. Contact him at tobias.wrigstad@it.uu.se.
https://orcid.org/0000-0002-4269-5408